\documentclass[sigconf]{acmart}

\AtBeginDocument{
  \providecommand\BibTeX{{
    \normalfont B\kern-0.5em{\scshape i\kern-0.25em b}\kern-0.8em\TeX}}}
\usepackage{array,booktabs,amsmath,graphics,fancyvrb,tabularx, longtable}
\usepackage[ruled,vlined,linesnumbered]{algorithm2e}
\usepackage{subfig}

\renewcommand\footnotetextcopyrightpermission[1]{}

\begin{document}

\title{\texttt{WattsApp}: Power-Aware Container Scheduling}

\author{Hemant Mehta}
\affiliation{\institution{University College Cork, Ireland}}
\email{h.mehta@cs.ucc.ie}

\author{Paul Harvey}
\affiliation{\institution{Rakuten Mobile, Japan}}
\email{paul.harvey@rakuten.com}

\author{Omer Rana}
\affiliation{\institution{Cardiff University, UK}}
\email{ranaof@cardiff.ac.uk}

\author{Rajkumar Buyya}
\affiliation{\institution{University of Melbourne, Australia}}
\email{rbuyya@unimelb.edu.au}

\author{Blesson Varghese}
\affiliation{\institution{Queen's University Belfast, UK}}
\email{b.varghese@qub.ac.uk}

\renewcommand{\shortauthors}{Mehta, et al.}

\begin{abstract}
  Containers are becoming a popular workload deployment mechanism in modern distributed systems. However, there are limited software-based methods (hardware-based methods are expensive requiring hardware level changes) for obtaining the power consumed by containers for facilitating power-aware container scheduling, an essential activity for efficient management of distributed systems. This paper presents \texttt{WattsApp}, a tool underpinned by a six step software-based method for power-aware container scheduling to minimize power cap violations on a server. The proposed method relies on a neural network-based power estimation model and a power capped container scheduling technique. Experimental studies are pursued in a lab-based environment on 10 benchmarks deployed on Intel and ARM processors. The results highlight that the power estimation model has negligible overheads for data collection - nearly 90\% of all data samples can be estimated with less than a 10\% error, and the Mean Absolute Percentage Error (MAPE) is less than 6\%. The power-aware scheduling of \texttt{WattsApp} is more effective than Intel's Running Power Average Limit (RAPL) based power capping for both single and multiple containers as it does not degrade the performance of all containers running on the server. The results confirm the feasibility of \texttt{WattsApp}.

\end{abstract}


\maketitle

\section{Introduction}
\label{sec:introduction}
Container technology is a lightweight virtualization technique that has low overheads when compared to Virtual Machines (VMs)~\cite{ruan2016performance}. Therefore, they are becoming popular for deploying workloads on clusters and clouds~\cite{hpc-container-1, hpc-container-2} and for upcoming distributed systems that use the edge of the network~\cite{challenge2016, wang2017enorm}. 

Container scheduling is an important avenue explored in the literature for distributed systems. Existing container scheduling strategies consider a number of relevant parameters, including resource demand, service level agreements and  hardware/software requirements~\cite{kaewkasi2017improvement, zhou2018scheduling}. However, container scheduling like any deployment strategy need to be power-aware so that the total power consumption of a system does not exceed predefined power cap limits. 

Modern processors are equipped with power capping techniques, such as Dynamic Voltage and Frequency Scaling (DVFS) and Running Average Power Limit (RAPL)~\cite{le2010dynamic, zhang2015quantitative}. These are hardware-based and reduce the CPU frequency and voltage to lower processor power consumption. However, this degrades the entire system performance and consequently the deployed application. 

It is valuable to gather the power consumption of individual containers running in a system so that they can be scheduled in a power-aware manner. However, there are limited software-based methods that measure container power consumption. Commercial vendors of Uninterrupted Power Supply (UPS) employ software-based power estimation techniques based on static information, such as input voltage for different types of devices (for example, laptops)\footnote{\url{https://www.apc.com/shop/uk/en/tools/ups_selector/}}, but ignore the workload-level granularity of power estimation. Other software based approaches, such as cWatts~\cite{phung2017application}, cWatts++~\cite{phung2019lightweight} and SmartWatts~\cite{Fieni2020} are either CPU architecture specific, do not capture all components of the system that contribute to container power consumption, and are intrusive methods (further considered in Section~\ref{sec:relatedwork}). 
This fundamental gap is addressed in this paper by developing \texttt{WattsApp}, underpinned by a six step software-based (not hardware-based since they are expensive and require hardware level changes), hardware architecture agnostic and relatively non-intrusive power-aware container scheduling method. The aim is to estimate power consumption of containers and schedule power capped containers to stay within safe power budgets. 

The research contributions of \texttt{WattsApp} are as follows:

(i)	A six-step software-based power-aware container scheduling method that accurately predicts power consumption of containers. The proposed method has negligible overheads (in relation to system power consumption) for collecting data required for estimating container power consumption. Additionally, nearly 90\% of all data samples can be estimated using the power model with less than a 10\% error. The Mean Absolute Percentage Error (MAPE) is observed to be between 1\%-6\%, which is relatively low. \texttt{WattsApp} is the first prototype that builds power models and enforces power capping for parallel applications that execute on a cluster of containers. 

(ii) The proposed power-aware method of \texttt{WattsApp} implements power capped scheduling for both single and multiple containers on the same server. The proposed power capping method is more beneficial than when no power cap or Intel's RAPL power cap is employed since the performance of all running containers on the system is not degraded (only containers that violate the budget are penalized). Potential approaches based on containers to achieve the power cap are to migrate the container to another server or deallocate resources of the container that violates the power cap. Experimentally, deallocating resources is a more viable approach than migration due to the inherent limitations of migrating containers. 

The remainder of the paper is organized as follows. 
Section~\ref{sec: Motivation} discusses the motivation for \texttt{WattsApp}. Section~\ref{sec:method} proposes the power-aware container scheduling method. The underlying power model is discussed in Section~\ref{sec:estimation}. Section~\ref{sec:scheduling} presents the power capped container scheduling approach. Section~\ref{sec:experimentalstudies} presents experimental studies. Section~\ref{sec:relatedwork} discusses the related work. Section~\ref{sec:conclusions} concludes the paper by considering future work.

\section{Background}
\label{sec: Motivation}

Predicting container power consumption is complex because it depends on the resource allocated to it and the workload running in the container. It is different when compared to the power prediction of VMs, other processes and hardware (processors, memory etc) because of the limited availability of data about the resource utilization and hardware performance counters specific to containers. This is because containers create multiple processes on the host operating system (OS). The number of processes varies depending on the activity that is performed within the containers. 

This paper observes that containers with more allocated resources consume more power for the same workload than on containers with fewer resources. However, in all cases the power consumed does not correspond to the increase in resources (CPU cores, memory). For example, the average power of an application running on twice the resources as another container, may not necessarily directly correspond to a factor of two.

\begin{table}[]
\centering
\caption{Scientific workloads used in this paper}
\label{table:workloads}
\begin{tabularx}{\linewidth}{l X c}
\hline
\textbf{Abbreviated Name} & \textbf{Description}                               & \textbf{Type} \\ \hline
KMEANS                    & Clustering algorithm                                          & DCBench                \\ \hline
FUZZY-KMEANS              & Clustering algorithm                                          & DCBench                \\ \hline
KPCA                      & Principal component analysis                                  & DCBench                \\ \hline
PCA                       & Principal component analysis                                  & DCBench                \\ \hline
BFS                       & Graph mining-breadth-first algorithm                          & DCBench                \\ \hline
MD                        & Molecular dynamics                                            & MPI-C          \\ \hline
HEATED                    & Steady heat equation solver                                   & MPI-C          \\ \hline
POISSON                   & Poisson equation solver in a rectangle using Jacobi iteration & MPI-C          \\ \hline
PRIME                     & Counting of prime in given limit                              & MPI-C          \\ \hline
SGEFA                     & Standard linear algebra solver                                & MPI-C          \\ \hline
\end{tabularx}
\end{table}

This hypothesis is verified on 10 different scientific workloads that are listed in Table~\ref{table:workloads}. These workloads are obtained from two sources. The first is DCBench, a bench marking suite~\cite{jia2013} from which five MPI based workloads are chosen. The second is a collection of C/C++ based scientific programs\footnote{\url{http://people.sc.fsu.edu/~jburkardt/c_src/c_src.html}}. These workloads are a combination of CPU-bound, I/O-bound and memory-bound scientific workloads that execute to completion. This paper does not consider alternate classes of workloads, such as Internet-of-Things, stream processing, or sensor-based applications. The workloads considered in this paper may have different power consumption patterns during execution. This is captured in the resource utilization and power data that is collected at a fine granularity and used for building the power model. This ensures that estimation can be carried out for different potential phases of a workload. 

\begin{figure}[]
  \centering
  \includegraphics [height=80pt, width=0.48\textwidth]{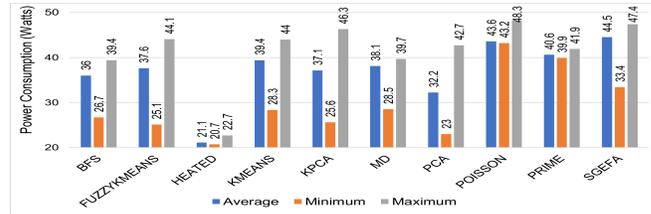}
  \caption{Power consumption (Watts) of workloads deployed in containers with 3 CPU Cores and 4GB RAM
  }
  \label{fig:powerstats1}
\end{figure}

\begin{figure}[]
  \centering
  \includegraphics [height=80pt, width=0.48\textwidth]{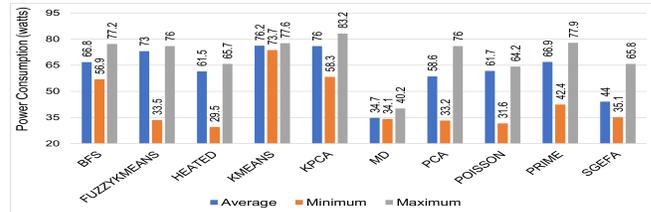}
  \caption{Power consumption (Watts) of workloads deployed in containers with 6 CPU Cores and 8GB RAM
  }
  \label{fig:powerstats2}
\end{figure}

Figure~\ref{fig:powerstats1} highlights the average, minimum and maximum power consumed by a container with 3 CPU cores and 4GB RAM.
 Figure~\ref{fig:powerstats2} provides results for the same workloads for twice the resources (6 CPU cores and 8GB RAM)
 It is evident that although the resources allocated are doubled the average power consumed does not necessarily double for all workloads (for example, refer to the workloads HEATED, MD, POISSON, PRIME and SGEFA).  

Similar trends are obtained when the power consumption is noted for the above workloads over time (the results are exhaustive and are not presented in this paper). When more resources are added parallel workloads (applications running within a single container) execute faster, but reach their peak power consumption at different times. This paper does not aim to explain individual power profiles of workloads, but notes that a server that executes multiple containers could violate the power cap; specifically, when multiple large size containers are multi-tenant on a server. These large containers may consume high power and when they are multi-tenant their combined total power consumption could be higher and close to the maximum power consumption of the server.

Power cap violations are undesirable and need to be effectively managed on servers running different workloads. They occur when the total power consumed by a server exceeds a threshold defined by the server administrators. When power cap violations occur, the server performance starts degrading since power management techniques like Dynamic Voltage and Frequency Scaling (DVFS) that are bundled with processors come in to play. DVFS reduces the server power consumption by using two power saving techniques, namely dynamic voltage scaling and dynamic frequency scaling~\cite{le2010dynamic}. Power saving is achieved by lowering the frequency and/or voltage of the CPU and other system resources. This reduction negatively impacts the performance of workloads executed on the server. For example, the performance of a container running on a server will drop when there is a power cap violation. 

In order to avoid the above, a power aware container scheduling strategy is required. It is observed that power/energy saving benefits of techniques like DVFS are diminishing because of the complexity in hardware technologies used for processors and memory (for example, increased memory performance and multi-core processors)~\cite{le2010dynamic}. Therefore, a software-based power capping technique is desirable in addition to specific hardware-based techniques. This motivates the power capping technique proposed in this paper.

\section{The \texttt{WattsApp} Method}
\label{sec:method}
This section presents a method for software-based power aware scheduling of containers to minimize power cap violations on a server, which is fundamental in developing \texttt{WattsApp}. \textit{Power aware container scheduling} is the distribution/consolidation of containers such that the total power consumed by a server does not cross a predefined threshold (or cap) specified by an administrator. 

A primary requirement for the \texttt{WattsApp} power aware container scheduling approach is obtaining information on the power consumed by an individual container. Resource utilization statistics of each container running on a server is used to calculate its power consumption. This information is used for container scheduling, such that the maximum power consumed does not violate any power restriction on an individual server. 

Currently, there are no hardware methods for obtaining the power consumed by containers. Moreover, there are a few software methods to measure the container power consumption and these methods have concerns like they are architecture specific, ignores essential system resources or intrusive as discussed in Section~\ref{sec:relatedwork}. This article aims to bridge this gap. Hardware-based methods will require modification of the hardware (such as additional probes) resulting in more expensive processors. Hence, a \textit{software-based method} is adopted to develop a model of container power consumption that depends on resource utilization information of the container. The model uses linear regression-based neural network to correlate container resource utilization statistics with system power consumption information to estimate the container power consumption. The model is further presented in Section~\ref{sec:scheduling}.

\begin{figure}[t]
  \centering
  \includegraphics[width=0.498\textwidth]{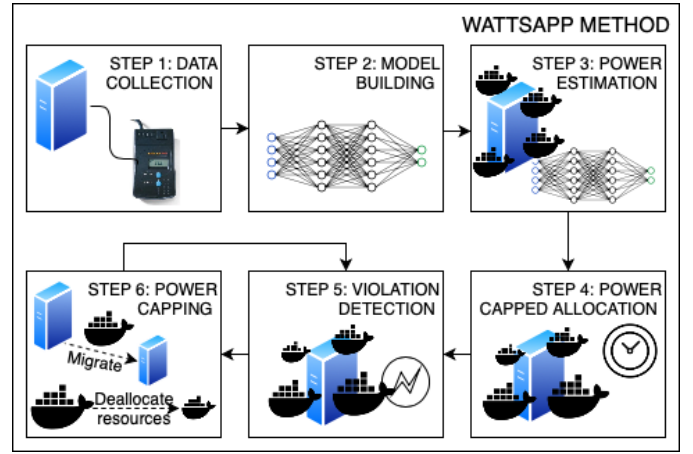}
  \caption{Six step method of \texttt{WattsApp} for power-aware container scheduling
  }
  \label{fig:method}
\end{figure}

The \texttt{WattsApp} method relies on two activities: 
(i) \textit{Container power prediction}, which is estimating the power consumed by an individual container using software-based methods, and 
(ii) \textit{Power capped container scheduling}, which is using the estimated power values to place containers equitably on a server. 

The importance of such a power-aware container scheduling method is that if the power cap exceeds on any server, then the performance of all containers running on the server will be degraded. To mitigate this, any power cap violation is periodically detected on each server by observing the total power consumption of the server. If the power consumed is above the threshold, then it is considered as a power cap violation. When a violation is detected, power capping is performed on the server without significantly affecting the run time performance of all running containers.

Power capping in \texttt{WattsApp} is performed using two approaches, namely container migration and container resource (CPU cores) reduction. The method aims to find a server on to which a container causing the server power capping violation can be migrated. If such a server is available, then the container is migrated to the identified server thereby avoiding any power cap violation.

If no servers are available to migrate a container, then a second approach is used by reducing the resources allocated to the container. The processor subsystem consumes nearly 85\% power of the total system power~\cite{mccullough2011evaluating}. Therefore, to achieve power capping the CPU cores allocated to the container are firstly reduced one at a time until the power cap limit is restored. 
\texttt{WattsApp} uses technology specific commands (such as \textit{docker update} command) to change the cores allocated to a container; the change is immediately reflected. As a result the container will use fewer CPU cores, which eventually reduces the server power consumption. 

When a container is migrated to a different server, the power consumption of the source server is again reduced below the power cap limit. A detailed discussion on both container migration and CPU core reduction is provided in Section~\ref{sec:scheduling}.

The proposed method of \texttt{WattsApp} shown in Figure~\ref{fig:method} comprises six steps: namely Data Collection, Model Building, Power Estimation, Power Capped Allocation, Violation Detection and Power Capping. The first three steps are for container power prediction, and the remaining steps are for power capped container scheduling.

\textbf{\textit{Step 1 - Data Collection}}: The training data for supervised learning is collected to estimate the power consumption of containers. It collects system power consumption and resource usage statistics for each container from the host OS. This is correlated with the system power consumption to obtain container power consumption. The system power consumption information is necessary as the regression techniques require labeled data for building a model. 

System power consumption data is collected from Watts Up .net hardware power meter. It facilitates the power consumption sampling at a one second granularity. Real time system power consumption data can be obtained by connecting it through a USB interface. It is reported that the accuracy of this hardware meter is +/- 1.5\%+0.3W\footnote{https://www.vernier.com/files/manuals/wu-pro.pdf}~\cite{hirst2013watts}. The readings provided are also considered to be generally reliable although a high error rate is observed for readings below 1~Watt~\cite{hirst2013watts}. A series of power meters from Watts Up are used in research reported in the literature~\cite{li2012power}\cite{appuswamy2013scale}\cite{bertran2012counter}\cite{asnaghi2016dockercap}.

\textbf{\textit{Step 2 - Model Building}}: The collected data is used to train data and build a neural network for individual containers running on the server. We build a model for each container on the server as different applications exhibits different properties. These models will be utilized during run time for predicting the power consumption of the container. 
The input to the neural network is container resource utilization statistics (including the percentage of host CPU and memory the container is using, the total memory the container is using and the maximum allocated memory, the amount of data the container sent and received on its network interface, and the amount of the data read from and written to the block input output devices) and system power consumption. The input data to the model is obtained in the previous step. 
The model (linear regression-based) was developed using the Keras\footnote{https://keras.io/} deep learning library.



\textbf{\textit{Step 3 - Power Estimation}}: Consider there are $n$ containers ($C_1, C_2, C_3 , \cdots C_n$) running on a server. The resource utilization statistics (CPU usage, memory usage, amount of block I/O and network data transfer) are collected for each container. The models developed (previous step) and the run-time statistics are used to predict the power consumed by the container.

\textbf{\textit{Step 4 - Power Capped Allocation}}: The models are used for power capped container scheduling. It uses the predicted power consumption of the container and current total power consumption of the server before scheduling the container on the server. The power consumption information of the containers are obtained by using the power estimation model that uses the power profile of each container for power capped allocation.


\textbf{\textit{Step 5 - Violation Detection}}: To perform power capped container scheduling a power cap violation has to be detected. After initial scheduling of a container, this step is executed at a five minute interval to check the server for any power cap violation. If there is a violation, then the final step enforces the power capping limit on the server.

\textbf{\textit{Step 6 - Power Capping}}: This final step adopts two techniques to enforce power capping. The first is referred to as migration - another server that can accommodate the container causing the power cap violation on a current server is identified; migration should not violate the power cap of the recipient server. If such a server is available, then the power cap violating container is migrated to the identified server. If no such server is identified, then a second technique, referred to as resource deallocation, is performed in which the number of CPU cores allocated to the container is reduced until the power cap limit is reached. This first prototype of \texttt{WattsApp} only considers a single container causing power cap violations. However, if multiple containers cause power cap violation, then a priority based container selection approach is required, which is not considered in this paper. 

The underlying approaches of the first three steps are presented in Section~\ref{sec:estimation}.
The last three steps are further discussed in Section~\ref{sec:scheduling}.

\section{\texttt{WattsApp} Container Power Estimation}
\label{sec:estimation}
This section describes the Data Collection (Step 1), Model Building (Step 2), and Power Estimation (Step 3) steps of \texttt{WattsApp}.

The \textbf{\textit{Data Collection}} step gathers (i) the system power consumption data, and (ii) resource utilization data of running containers. This is a black box technique as the data is collected from the host operating system and no profiling data is obtained from within the container. The other approaches (referred to as white box) collect profiling information inside the container and should be avoided to maintain the integrity of the containers~\cite{gu2014power}.  

Power consumption data is collected from the Watts Up .net power meter. The data obtained contains the time stamp and power consumption (in Watts). The resource utilization data of the container is collected using  the \texttt{docker stats} command, which provides the following data: (i) Id of the container and the name of the container, (ii) Percentage of host CPU and memory the container is using, (iii) Total memory the container is using and the maximum allotted memory, (iv) The amount of data the container has sent and received on its network interface, (v) The amount of the data read from and written to the block input/output devices, and (vi) The number of processes/threads created by container.

The data collection time stamp is also added to the output of docker stats. Both the power consumption data and the resource utilization data of the container is concatenated with respect to the time stamp. 
Multiple CPU cores allocated to the container are taken care of by the host CPU usage (for example, it ranges from 0 to 100\% for one core, up to 200\% for two cores and so on.

Resource utilization and system power consumption data are obtained once per second during the execution of the benchmarks. The sequence of steps is presented in Algorithm~\ref{alg:datacollection}.

\begin{algorithm}[h]
\DontPrintSemicolon
\SetAlgoLined
\SetKwInOut{Input}{Input} \SetKwInOut{Output}{Output}
\Input{Container name}
\Output{Combined resource utilization and power consumption data}
\BlankLine

\While {workload in container is running}{
    Obtain resource utilization from docker stat command once per second\;
	Add time stamp to the output of docker stat\;
	Collect Watts Up power data once per second\;
}
Combine resource utilization and power data on the basis of the timestamp \;
\caption{Data Collection}
\label{alg:datacollection}
\end{algorithm}

In the \textbf{\textit{Model Building}} step, a regression technique that relies on a neural network is used. 
The inputs to the model are the container resource utilization (including percentage of host CPU and memory the container is using, total memory the container is using, maximum allotted memory, amount of data the container has sent and received on its network interface, amount of the data read from and written to the block input output devices), and system power consumption. The output is container power consumption. 

In the \textbf{\textit{Power Estimation}} step, the power consumption of containers is firstly modeled by \texttt{WattsApp}. The power consumption of a server ($P_{server}$) comprises static power and dynamic power. Static power ($P_{static}$) is defined as the power consumption of system when there is no active container. This power is measured by using the Watts Up .net power meter. If there is only one running container then the total power consumption is sum of idle power consumption and dynamic power consumption of the server. The dynamic power consumption ($P_{dynamic}$) is defined as the power consumption of the running container.
\begin{equation}
P_{server} = P_{static} +  P_{dynamic} 
\label{server_model_short}
\end{equation}

If there are $n$ active containers on the server, then the dynamic power consumption of the server is the aggregate power consumption of all the containers.
\begin{equation}
 P_{dynamic} = \sum_{k=1}^{n} P_{container_{k}} 
\label{dynamic_power}
\end{equation}
where $P_{container}$ is the power consumption of the container.

The dynamic power consumption of the system is considered as the sum of the power consumed by the CPU, the memory (RAM), the disk and the network.
\begin{equation}
P_{container_{k}} =  a_{k}*Ucpu_{k} +  b_{k}*Uram_{k}  +  c_{k}*Udisk_{k} +  d_{k}*Unet_{k}    
\label{vm_model}
\end{equation}
where $a, b, c,$ and $d$ are constants, $n$ is the number of running containers in a server, $l$ represents the number of CPU cores allocated to a container, $Ucpu$ is the CPU utilization factor, $Uram$ is the RAM utilization factor, $Udisk$ is the disk utilization factor, and $Unet$ is the network utilization factor.

\begin{equation}
Ucpu =  \sum_{i=1}^{l} Ucpu_{{core}_{i}}    
\label{vm_model}
\end{equation}

\begin{multline}
P_{server} = P_{static} + \sum_{k=1}^{n} a_{k}*Ucpu_{k} + \sum_{k=1}^{n} b_{k}*Uram_{k} \\ + \sum_{k=1}^{n} c_{k}*Udisk_{k} + \sum_{k=1}^{n} d_{k}*Unet_{k}  
\label{server_model_exp}
\end{multline}

The case when a single workload is executed across a cluster of containers is considered by \texttt{WattsApp}. In this case, the workload's power consumption will be the aggregate power consumption of all the containers of the cluster.

\begin{equation}
 P_{workload} = \sum_{i=1}^{n} P_{Container_{k}} 
\label{application_power}
\end{equation}
where $Container_{k}$ is element of $C$, the set of containers in the cluster $C= (Container_{1}, Container_{2}, \cdots Container_{n})$. 


\section{\texttt{WattsApp} Power Capped Container Scheduling}
\label{sec:scheduling}
This section presents the use of the estimated power values at run time for a proposed power capped container scheduling method of \texttt{WattsApp}. 
The approach adopted is to initially allocate containers using a best fit or first fit strategy, and subsequently when there is a power cap violation on the server, migrate the container elsewhere or reduce the allocated CPU cores of the running containers until the power capping is not violated. 
Power Capped Allocation (Step 4), Violation Detection (Step 5), and Power Capping (Step 6) proposed in Section~\ref{sec:method} is considered in this section.

\textbf{\textit{Power Capped Allocation}} uses the estimated power consumption of containers to schedule containers by calculating the total power consumption of the candidate server after adding the estimated power consumption of the container ready for deployment. If the power consumption of the server is anticipated to be below the power cap limit, then the container will be placed on the candidate server. This is repeated for all containers that are ready for placement. Algorithm~\ref{alg:powercapplacement} highlights this and Table~\ref{table:notation} presents the notation used in this algorithm and the other algorithms (Algorithm~\ref{alg:detectpowercap} and Algorithm~\ref{alg:applypowercap}) presented in this section. It is assumed that there are $n$ servers, and each server may have up to $m$ containers. 

\begin{algorithm}[t]
\DontPrintSemicolon
\SetAlgoLined
\SetKwInOut{Input}{Input}
\Input{$S$, $C_{i_j}$, $md$}
\BlankLine

\For {$\forall$ $c_i$ $\varepsilon$ $C$} {
    $flag=false$ \;
    \For {$\forall$  $s_i$ $\varepsilon$ $S_i$} {
        $PS_i=PS_i+containerPower_i$ \;
        \eIf {$serverPower_i$ $<$ $cap$} {
            allocate($c_i$, $s_i$) \;
            $flag=true$
        } {
            do nothing\;
        }
    }
    \If {$flag == false$} {
       Power Capped allocation is not possible \;
       \BlankLine
       Select the $i^{th}$ server with minimum $PS_i$ to place the current container
    }
}
\caption{Power Capped Scheduling of Containers}
\label{alg:powercapplacement}
\end{algorithm}

\begin{table}[t]
\centering
\caption{Notation used in the Power Capping Method}
\label{table:notation}
\begin{tabularx}{\linewidth}{c p{6.8cm}}
\hline
\textbf{Notation}                                   & \textbf{Description}\\\hline
$S$                                                 & Set of all the servers $s_i$ $\varepsilon$ S for $i = 1, 2, \cdots, n$                                                                            \\\hline
$Cr_i$                                              & Number of CPU Cores in server $s_i$             \\\hline

$M_i$                                               & Available memory in the server $s_i$             \\\hline
$C_{i_j}$                                           & List of containers deployed in server  $S_i$ for   $j = 1, 2, \cdots, m$ \\\hline
$C$                                                 & List of all the containers ready to be placed   
    \\ \hline
$ACr_i$                                             & List of CPU core allocation to each container    \\ \hline
$AM_i$                                              & List of memory allocation to each container       \\ \hline
$md$                                                & Power consumption Model           \\ \hline
$PC_{i_j}$                                          & Power consumption of container $c_{i_j}$            \\ \hline
$PS_i$                                              & Calculated total power consumption of server $s_i$,  $ps_i$ $\varepsilon$ PS                                                     \\\hline
$c_j$, $pc_j$                                             & The candidate container and its power consumption causing the server power increasing beyond the power cap   threshold \\\hline
cap                                                 & Power cap                               \\                      \hline                                                                                
\end{tabularx}
\end{table}

Algorithm~\ref{alg:powercapplacement} executes for all containers ready for placement (line 1). The flag variable is initialized to false (line 2); this variable will be used to identify the case when no suitable server for power capped placement. Each server is checked one by one whether it can accommodate the container under consideration (line 3). The container power consumption is added to the candidate server power (line 4) to check if it can accommodate the container (line 5). If the candidate server can deploy the container, then it is allocated to the server (line 6). When container placement is successful, the flag is updated to true (line 7). If it is not possible to allocate on a given server, then the remaining servers are processed. When no suitable server is found for power capped placement (line 12), the container is allocated to the server with the lowest power consumption (line 13 and 14). After this, the  Algorithm~\ref{alg:detectpowercap} will work to detect the possibility of power cap violation, and if required, the power cap is applied using Algorithm~\ref{alg:applypowercap}.  


A process to determine any power cap violation is executed on the servers, five minutes after initially scheduling containers (profiling data is collected for the first five minutes). This process uses the power prediction model to estimate the power consumption of each running container on the server. If the power consumption of any server (sum of power consumption of all the running containers) is beyond the power cap, then there is a power cap violation caused by the newly placed container). This container will be considered as the candidate for migration or CPU core reduction.  

Power capping is achieved in two ways. The first is by \textit{migrating the candidate container} from the source to a destination server whose current power consumption is below the cap and would not be violated if it accepted the container. A stateful migration method, namely `CRIU (Check-point/Restore In Userspace)' is employed for migrating containers.



The second is by reducing the resources allocated to the candidate container, specifically the number of CPU cores (reduce one at a time) as CPU usage significantly affects the power consumption of Docker containers~\cite{tadesse2017energy}. The Docker update command is used to change the number of allotted CPU cores to the container. 
The performance of the container will be degraded when using this approach (further considered in Section~\ref{sec:experimentalstudies}). 

Currently, there is support for power capping on the hardware. However, most hardware power capping techniques tweak the voltage and processor frequency, which affects the potential performance of the entire system and is detrimental to \textbf{all} containers running on the system \cite{le2010dynamic}. However, the proposed software power capping technique achieves the power cap without significantly affecting the entire system's performance and only negatively impacts the container that causes the power cap violation. 

\textbf{\textit{Violation Detection}} is to detect when a power cap violation occurs on any server under consideration. The detection algorithm is given in Algorithm~\ref{alg:detectpowercap} and runs on each of the servers. When a violation is detected, Algorithm \ref{alg:applypowercap} performs power capping. The model used to compute the power consumption of the container at run-time is considered in Section~\ref{sec:estimation}. 


\begin{algorithm}[t]
\DontPrintSemicolon
\SetAlgoLined
\SetKwInOut{Input}{Input}
\Input{$S$, $C_{i_j}$, $md$}
\BlankLine

\For {$\forall$ $s_i$ $\varepsilon$ $S$} {
    $totalPower_i=0$ \;
    \For {$\forall$ $c_j$ $\varepsilon$ $C_{i_j}$}{
        data = collect\_stats($c_j$) \;
        $PC_{i_j}$ = md.predict(data)\;
        $totalPower_i = totalPower_i + PC_{i_j}$\;
    }
}
\For {$\forall$  $s_i$ $\varepsilon$ $S_i$} {
    \eIf {$totalPower_i$ $>$ $cap$} {
        powerCap($c_j$, $ps_j$,$s_i$, $PS$) \;
    } {
        do nothing\;
    }
}
\caption{Power Cap Violation Detection}
\label{alg:detectpowercap}
\end{algorithm}

The detection algorithm firstly computes the power consumption of each server (line 1) indirectly by calculating the power consumption of every container (line 3) deployed on the server. This is achieved by collecting the resource usage statistics for each container (line 4) and then passing the data to the power model for predicting power consumed (line 5). The power consumption of all containers running on a server are summed to obtain server power consumption (line 6) after which power cap violation (if any) is checked for on each individual server (line 9 and 10). If a server crosses the power cap limit then Algorithm~\ref{alg:applypowercap} is performed with the required inputs. If the power consumed by all containers on the servers is below power cap, then no changes are made (line 13).

The last step is \textbf{\textit{Power Capping}} that uses Algorithm~\ref{alg:applypowercap} and the input provided by Algorithm~\ref{alg:detectpowercap}.
A server (line 1) that can accommodate the container without crossing the power cap (line 2) is searched for by checking with a network process that runs on each server. If successful, then the container that violates the power cap on a source server is migrated to the destination server (line 3). If no candidate destination servers are found, the algorithm uses the second option of reducing the allocated CPU cores to a container until the server power consumption falls below the power cap (line 7). For this, the allocated cores are first reduced by 1 (line 8) and then the resource usage statistics are collected (line 9) and the power consumption is predicted (line 10). Again, if the current power consumption falls below the power cap (line 11) then the algorithm is successful and returns (line 12). 

\begin{algorithm}[t]
\DontPrintSemicolon
\SetAlgoLined
\SetKwInOut{Input}{Input} \SetKwInOut{Output}{Output}
\Input{$S$, $C_{i_j}$, $c_{j}$, $ACr_{i}$, $cap$, $PS$, $md$}
\Output {true if power capping is successful, false otherwise}
\BlankLine

\For {$\forall s_{j}$ $\varepsilon$ ($S$ - $s_{i}$)} {
    \If {$PS_{j}$ + $pc_{j}$ $<$ $cap$} {
                migrate($c_{j}$, $s_{j}$) \;
               return true \;
    } 
}
\While {$ps_{j} > cap$}{
       reduceCoresByOne($c_{j}$) \;
       data = collectStats($c_{j}$) \;
       predictedpower=md.predict(data) \;
       \If{$ps_{j} < cap$}{
            return true \;
        }
}
return false \;
\caption{Apply Power Capping}
\label{alg:applypowercap}
\end{algorithm}

CPU core reduction to achieve power capping will degrade performance of the selected container. This can be compensated for by increasing the CPU cores at a later stage when it may be feasible to do so without exceeding the power cap. This scenario is considered in experimental studies to demonstrate that the impact of CPU core reduction of a container can be compensated when running the parallel component of an application in a cluster.

There may be a delay in enforcing the power capping limit since the detection algorithm is only executed once every five minutes (this time is a configurable parameter of the algorithm to suit any bespoke requirements). Experimentation on the impact of this limit is not presented in this article. It was empirically observed that after a power cap violation was detected, a few minutes were required for migrating the container to another server. The time taken for migration is due to: (i) Creating checkpoints - a checkpoint freezes a running container and turns its state into a collection of files on disk, (ii) Compressing the checkpoint and transferring it to the selected server, and (iii) Creating a new container and restoring the checkpoint. The majority of the migration time (depends on the size of the container image) is for transferring the checkpoint to the destination server. 

\section{Performance Evaluation}
\label{sec:experimentalstudies}

The experiments highlight that for \texttt{WattsApp}: (i) the overheads for data collection to estimate power do not significantly impact system power consumption, (ii) there is limited error in estimating power using the model, (iii) \texttt{WattsApp} operates across multiple processors, and (iv) the proposed power capping method is more effective for scheduling than alternate methods, such as Intel's RAPL. 

The experiments are conducted on two systems with different form factors. The first is workstation Dell Precision 3630 with an Intel Xeon E-2174G processor and 16GB memory. The second is a small form factor Odroid N2 Board with a quad-core ARM Cortex-A73 CPU cluster and a dual core Cortex-A53 cluster and 4GB memory. The system power consumption data is collected using Watts Up .net hardware power meter. The systems run on Ubuntu 18.04. The containers are created with Ubuntu 18.04 LTS image. Each container is allocated three CPU cores, 4GB memory on workstation and 2 CPU cores and 2GB memory on Odroid. Docker 17.12.0-ce version is used to deploy the containers. Keras 2.2.0 that runs on TensorFlow 1.8.0 is used to build the power model. The hardware power meter Watts Up .net is used to obtain system power consumption in real-time. Watts Up .net power is used to collect the power readings using USB from the host OS by a Python script that reads instantaneous power data.

The workloads defined in Table~\ref{table:workloads} are used for evaluating the \texttt{WattsApp} method. These are scientific workloads that execute to completion. This paper does not consider alternate workloads, such as Internet-of-Things, stream processing, or sensor-based applications. The experiments are carried out for single workloads on single containers, multiple workloads on multiple containers, and single workload across multiple containers (a cluster of containers) to thoroughly evaluate the \texttt{WattsApp} method. 

\begin{figure}[t]
  \centering
  \includegraphics[width=0.48\textwidth]{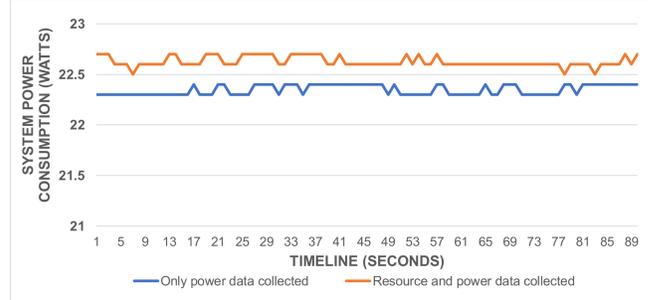}
  \caption{Data collection overhead for an 89 second time period on Intel Xeon processor}
  \label{fig:intel-data-overhead}
\end{figure}

\begin{figure}[t]
  \centering
  \includegraphics[width=0.48\textwidth]{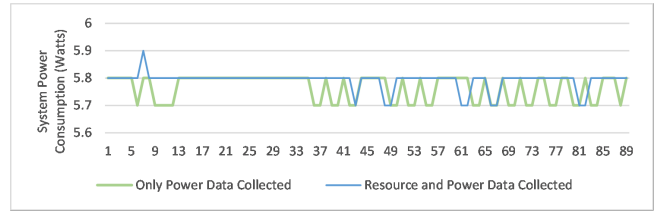}
  \caption{Data collection overhead for an 89 second time period on ARM processor}
  \label{fig:arm-data-overhead}
\end{figure}

\textbf{\textit{Results:}}
The data collection overheads and estimation error in container power prediction is firstly presented. Then, the results from scheduling for power capping obtained for single and multiple containers are considered. 
The average system power consumption overhead when collecting data (CPU, memory, and disk utilization along with power) for a 89 second time period is shown in Figure~\ref{fig:intel-data-overhead}. The blue plot shows the average system power consumption when only power data is collected, and the orange plot shows when both resource utilization and power metrics are collected. On an average, nearly 0.2~Watts are spent. The graph illustrates that the overhead in terms of system power is negligible (less than 1\% of system power consumption). This is an indicator that Step~1 of the proposed power-aware scheduling method is feasible. Figure~\ref{fig:arm-data-overhead} shows the data collection overhead on the ARM processor; the overhead is nearly 1.7\% of the system power consumption. 

\begin{figure}[t]
  \centering
  \includegraphics[width=0.48\textwidth]{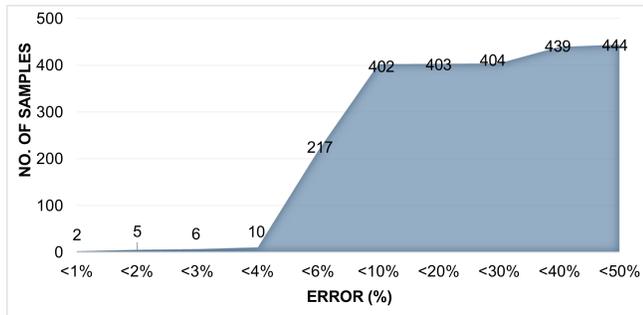}
  \caption{Error distribution for 444 samples of data on the Intel Xeon processor}
  \label{fig:intel-error-distribution}
\end{figure}

Figure~\ref{fig:intel-error-distribution} shows the distribution of error on the Intel Xeon processor in the power values that is estimated for 444 samples using the neural network model. In this experiment, data collected from all workloads (Table~\ref{table:workloads}) is consolidated to build the model that is validated using repeated random sampling by splitting data into 75\% and 25\% as training and testing dataset respectively. More than 90\% of the samples have an error of less than 10\% and nearly 49\% of the samples have less than a 6\% error. This highlights that the power model built in Step 2 of the method will have a reasonable accuracy for prediction in Step 3.    

\begin{figure}[t]
  \centering
  \includegraphics[width=0.48\textwidth]{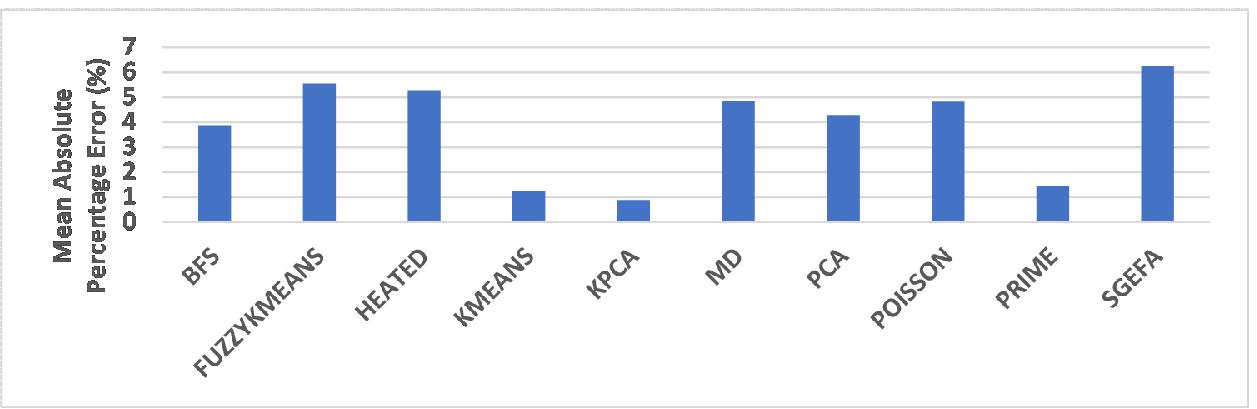}
  \caption{Mean Absolute Percentage Error (MAPE) in estimating system power for different workloads running within containers on the Intel Xeon processor}
  \label{fig:intel-mape}
\end{figure}

Figure~\ref{fig:intel-mape} highlights the Mean Absolute Percentage Error (MAPE) of the model for estimating power of individual workloads executing in a container on the Intel Xeon processor. For this experiment each workload is executed in a single container and the power is estimated for each container. MAPE indicates the average of percentage errors (a lower value indicates that the model estimates the power consumed with a higher accuracy). The average percentage error is between 1\% and just over 6\%. 

\begin{figure}[t]
  \centering
  \includegraphics[width=0.48\textwidth]{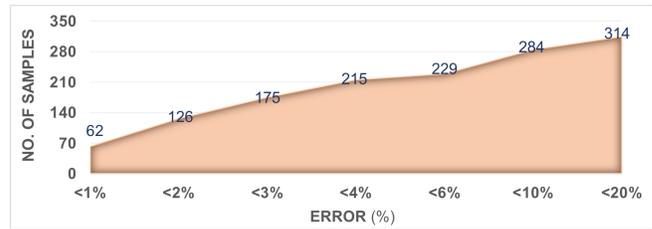}
  \caption{Error distribution for 400 samples of data on the ARM processor}
  \label{fig:arm-error-distribution}
\end{figure}

Similar experiments are performed on the Odroid board with ARM processor using six workloads from (Table~\ref{table:workloads}) (the four workloads, namely FUZZYKMEANS, KMEANS, KPCA and PCA, are distributed in DCBench with the binaries for the x86 platform). 
Figure~\ref{fig:arm-error-distribution} shows the error distribution of estimating power values for 400 samples using the neural network model. In this experiment, data collected from the six workloads is consolidated to build the model that is validated using repeated random sampling by splitting data into 75\% and 25\% as training and testing dataset, respectively. More than 80\% of the samples have an error of less than 15\% and more than 60\% of the samples have less than a 10\% error. This highlights that the power model from Step 2 will have a reasonable prediction accuracy in Step 3.    

\begin{figure}[t]
  \centering
  \includegraphics[width=0.48\textwidth]{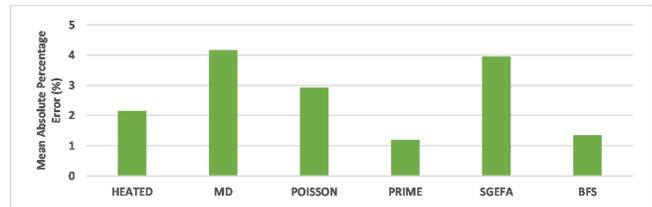}
  \caption{Mean Absolute Percentage Error (MAPE) in estimating system power for different workloads running within containers on the ARM processor}
  \label{fig:arm-mape}
\end{figure}

Figure~\ref{fig:arm-mape} highlights the Mean Absolute Percentage Error (MAPE) of the model for estimating power of individual workloads executing in a container on the ARM processor. For this experiment each workload is executed in a single container and the power is estimated for each container. MAPE indicates the average of percentage errors (a lower value indicates that the model estimates the power consumed with a higher accuracy). The average percentage error is between 1\% and just over 4\%. 

\begin{figure}[t]
  \centering
  \includegraphics[height=80pt, width=0.48\textwidth]{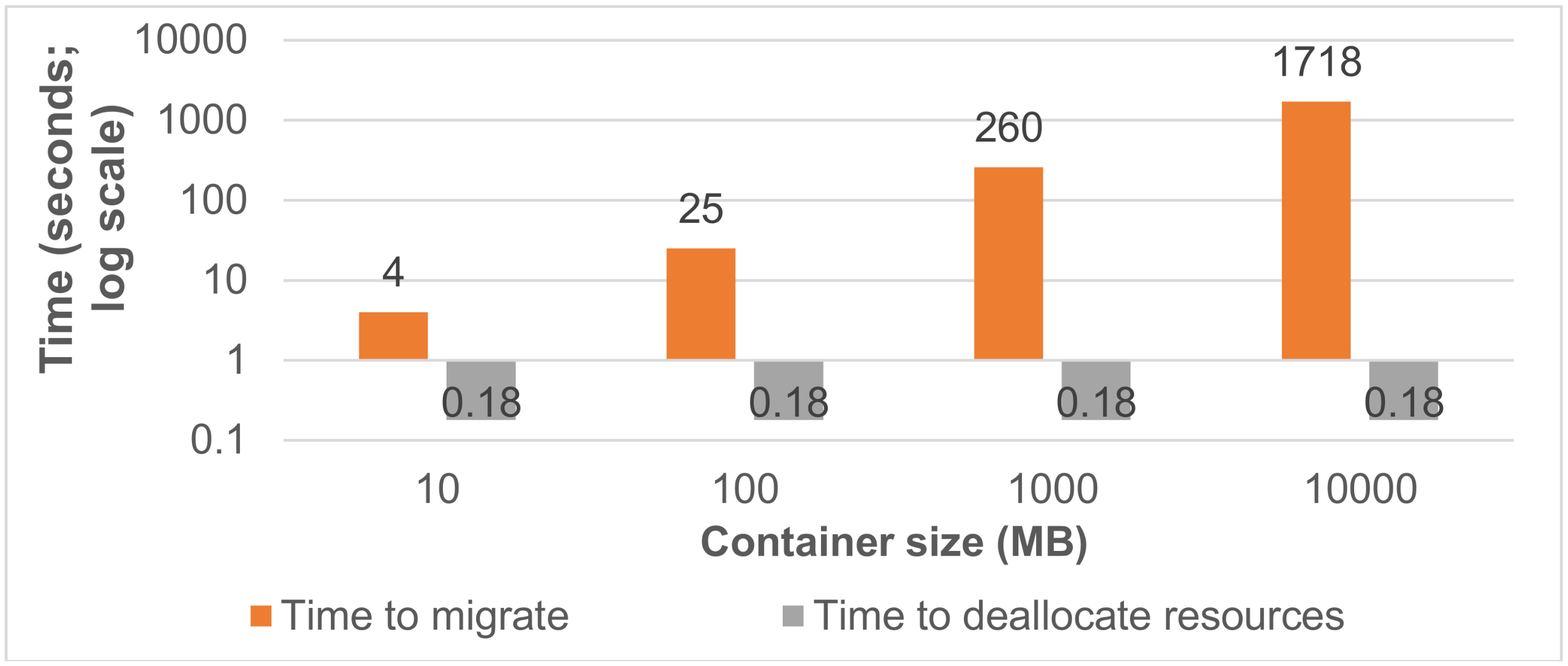}
  \caption{Time taken for migrating containers and deallocating resources of a container in the power capping method for containers of different sizes}
  \label{fig:overhead-powercapping}
\end{figure}

Figure~\ref{fig:overhead-powercapping} shows the overheads associated with the two techniques adopted in power capping, namely migration and deallocation of resources for different sizes of containers. The time taken to migrate using the Checkpoint/Restore in Userspace approach provided by Docker is directly proportional to the size of the container as the container needs to be checkpointed and migrated to an alternate server. However, using the time taken to deallocate resources on a container on the server takes approximately 180 milliseconds. Although migration is a potential option to achieve the power cap, the results show that deallocating resources is a more viable option given the inherent overheads in container migration. In the next set of experiments, power capping results based on only resource deallocation is presented. Migration using containers is a less viable option based on existing technology (if a critical application has to be executed) given large migration overheads although it may be lower than VMs (also not suited for single parallel application executed across a cluster of containers).


\begin{figure}[t]
  \centering
  \includegraphics[height=70pt, width=0.48\textwidth]{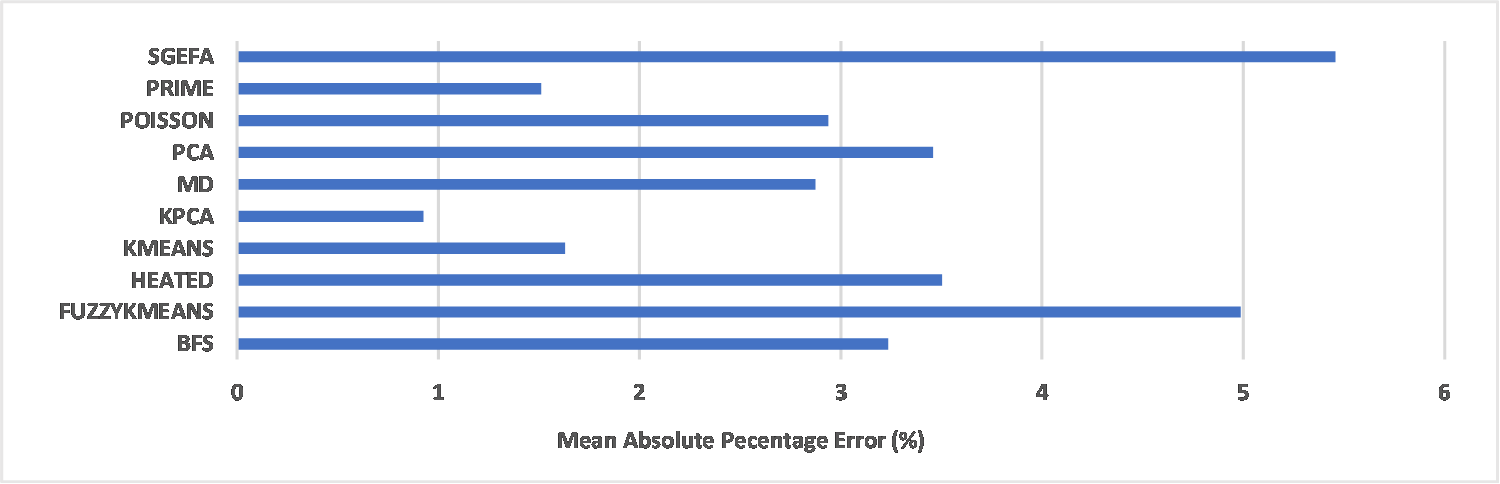}
  \caption{MAPE of estimating system power for parallel workloads running across a cluster of four containers on Intel Xeon processor}
  \label{fig:cluster-mape}
\end{figure}

\begin{figure}[t]
  \centering
  \includegraphics[height=70pt, width=0.48\textwidth]{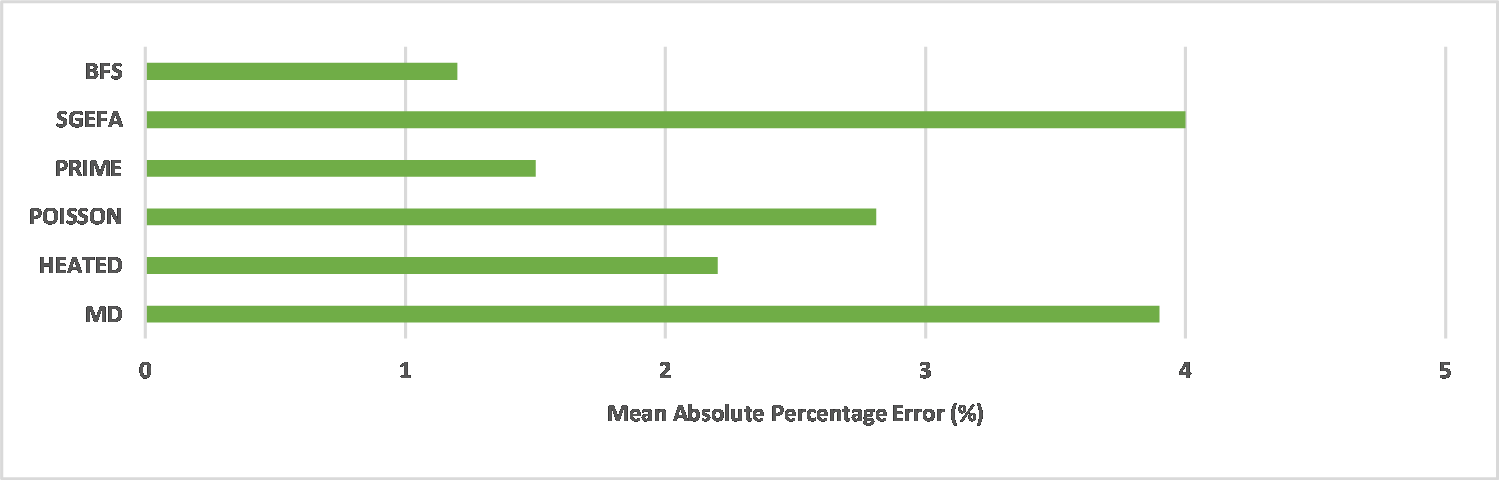}
  \caption{MAPE of estimating system power for parallel workloads running across a cluster of two containers on ARM processor}
  \label{fig:cluster-mape-arm}
\end{figure}

In another experiment, a cluster of containers (four on the Intel processor and two on ARM) was created for running the MPI applications from Table~\ref{table:workloads}. Figure~\ref{fig:cluster-mape} shows the results on the Intel processor to demonstrate the feasibility of container power prediction for parallel applications executed across multiple containers.
The average MAPE is 3 with error between 1 \% and around 5.5\%. Figure~\ref{fig:cluster-mape-arm} show the MAPE on ARM processor. The average MAPE 2.6 with error between 1 \% and 4 \%.

\begin{figure}[t]
\begin{center}
	\subfloat[BFS] 
	{\label{sfig:bfs}
	\includegraphics[width=0.142\textwidth]
	{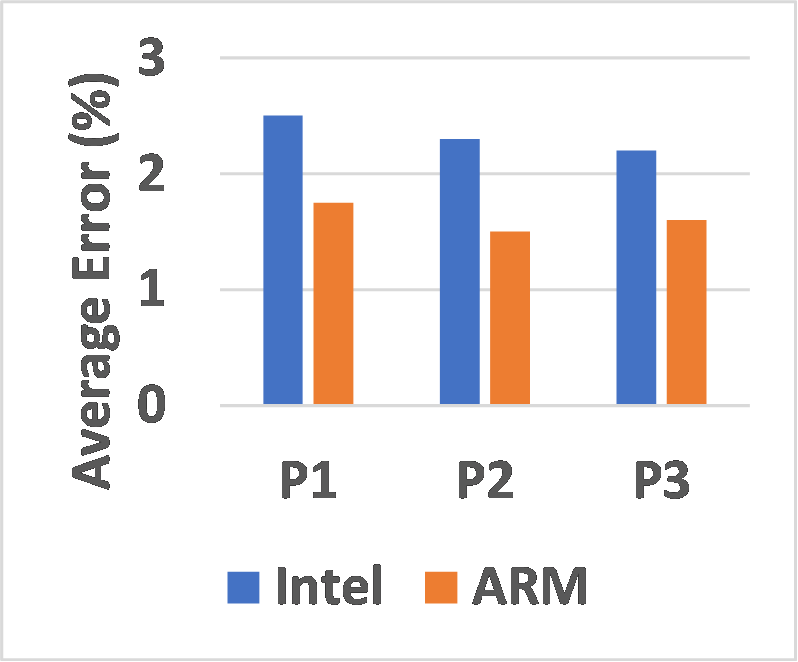}}
    \quad
	\subfloat[POISSON] 
	{\label{sfig:poisson}
	\includegraphics[width=0.142\textwidth]
	{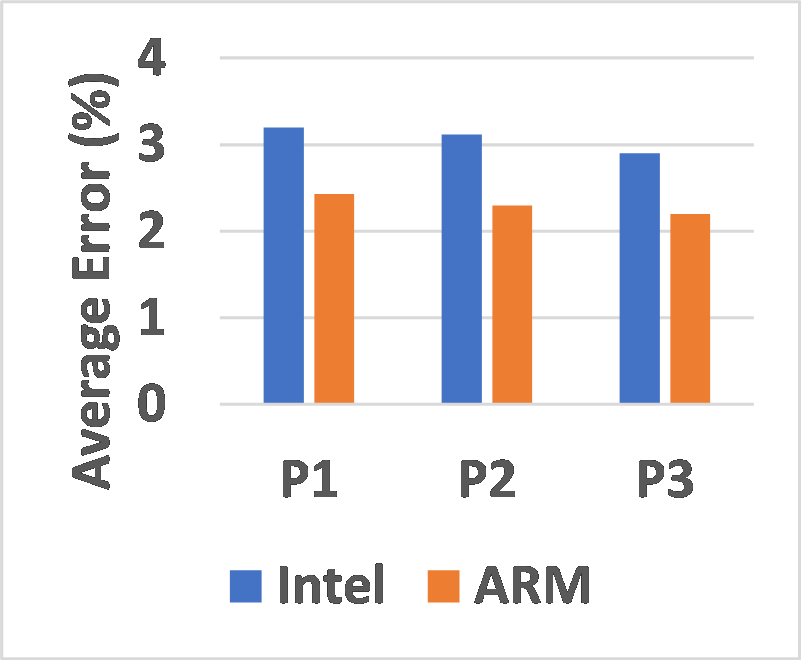}}
    \quad
	\subfloat[MD]
	{\label{sfig:md}
	\includegraphics[width=0.142\textwidth]
	{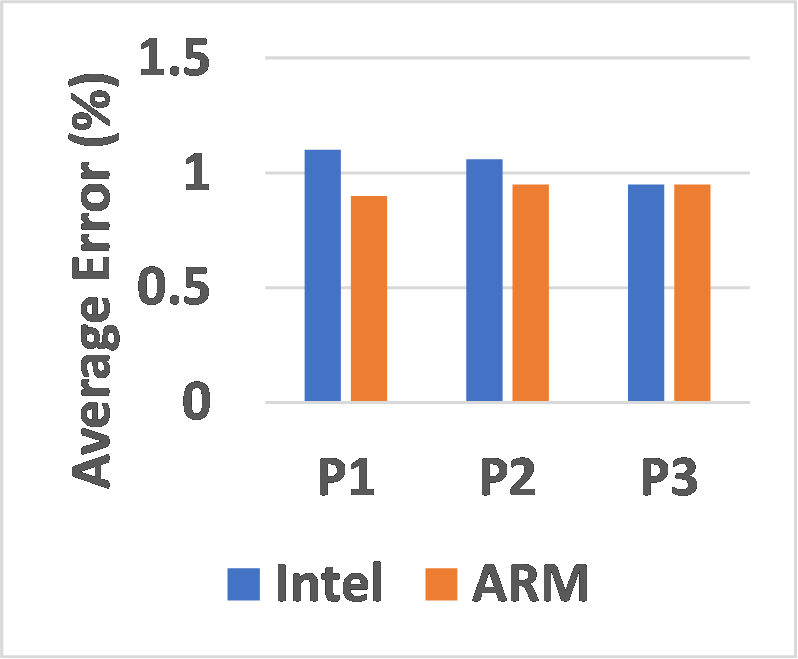}}

\end{center}
\caption{Average percentage error in the measured versus estimated power consumption using \texttt{WattsApp} for three different input configuration values of selected benchmarks running on a single container}
\label{fig:sensitivity}
\end{figure}

\begin{figure}[t]
\begin{center}
	\subfloat[BFS] 
	{\label{sfig:bfscluster}
	\includegraphics[width=0.142\textwidth]
	{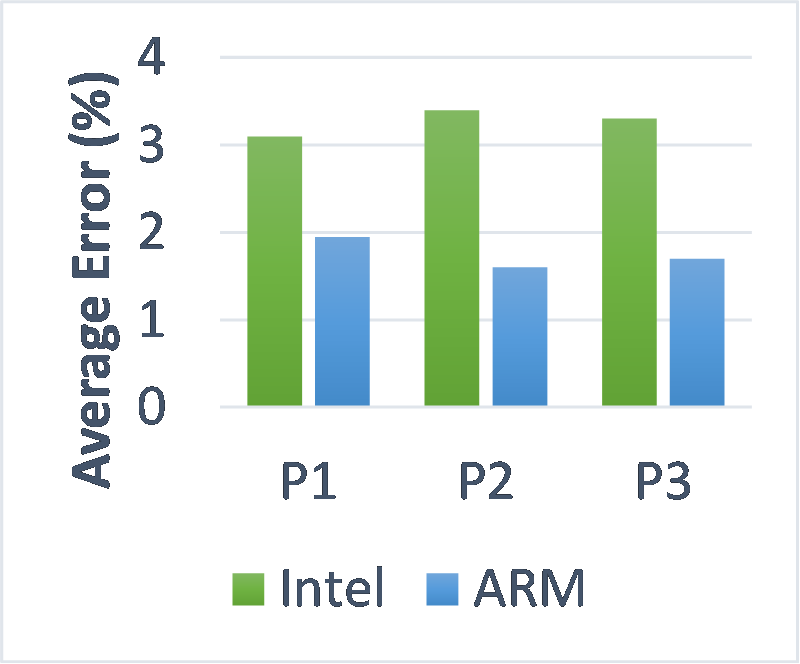}}
    \quad
	\subfloat[POISSON] 
	{\label{sfig:poissoncluster}
	\includegraphics[width=0.142\textwidth]
	{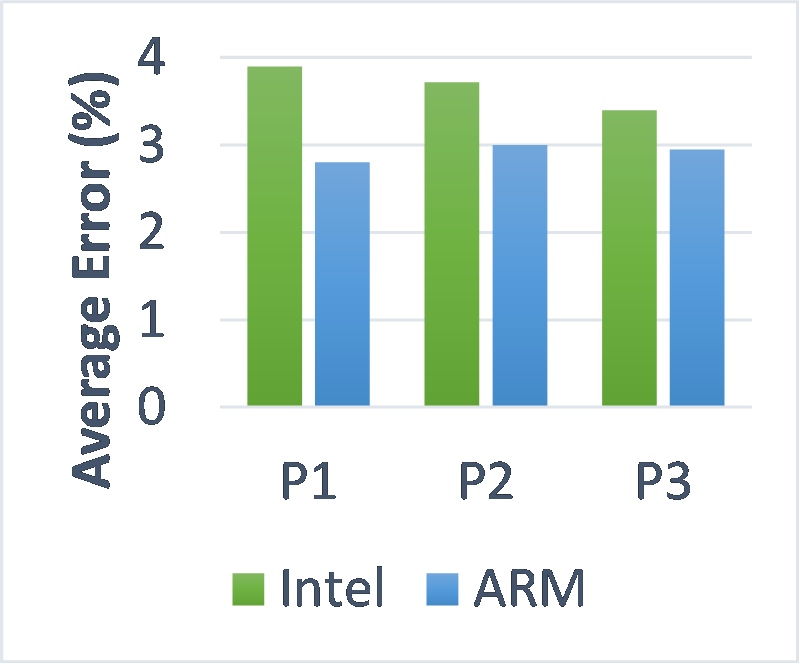}}
    \quad
	\subfloat[MD]
	{\label{sfig:mdcluster}
	\includegraphics[width=0.142\textwidth]
	{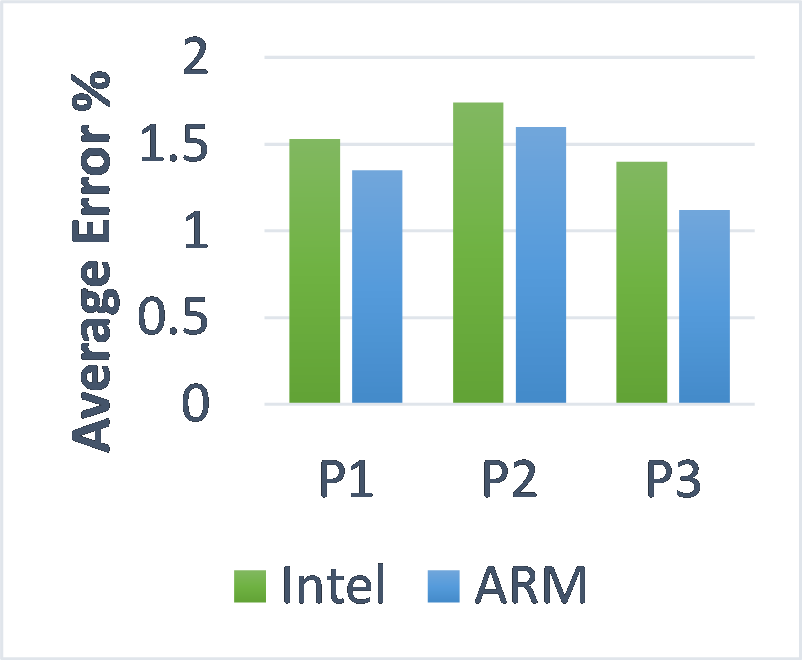}}

\end{center}
\caption{Average percentage error in the measured versus estimated power consumption using \texttt{WattsApp} for three different input configuration values of selected benchmarks running across two containers in a cluster}
\label{fig:sensitivitycluster}
\end{figure}

The accuracy of \texttt{WattsApp} power estimation is considered for different input parameters when the benchmark is executed in a single container (Figure~\ref{fig:sensitivity}) and in cluster of two containers (Figure~\ref{fig:sensitivitycluster}). Only three benchmarks (BFS, POISSON, MD) are presented with three different input parameters (P1, P2 and P3). The input to BFS is the parameter scale for which P1, P2, and P3 values are 8, 12, and 16 respectively. POISSON takes as input the number of interior vertices in one dimension, for which we chose P1, P2, and P3 as 16, 32, and 64 respectively. MD requires parameters: spatial dimension, number of particles, number of time steps and time step size; P1 = \{2, 500, 500, 0.2\}, P2 = \{3, 500, 500, 0.2\}, and P3 = \{3, 750, 500, 0.2\}. The data for the input parameters were not used during training. The results highlight that the average error percentage is between 0.8\% and 4\% for both Intel and ARM processors.

Two further experiments were carried out on the Intel processor to identify the potential benefit of the proposed power capping based container scheduling compared against when no power caps and Intel's RAPL-based power cap is employed. The first experiment is when a single container executes on the server with a given workload. In this experiment, there is only one container running on the server that is likely to violate the power cap. The second experiment is when multiple (three) containers that run the same workload execute on the server. There are multiple containers running on a given server, and any one of them may violate the power cap. Each container executes the same workload.

\begin{figure}[t]
\begin{center}
	\subfloat[For DCBench programs from Table~\ref{table:workloads}]
	{\label{fig:execution-time-single-1}
	\includegraphics[width=0.48\textwidth]
	{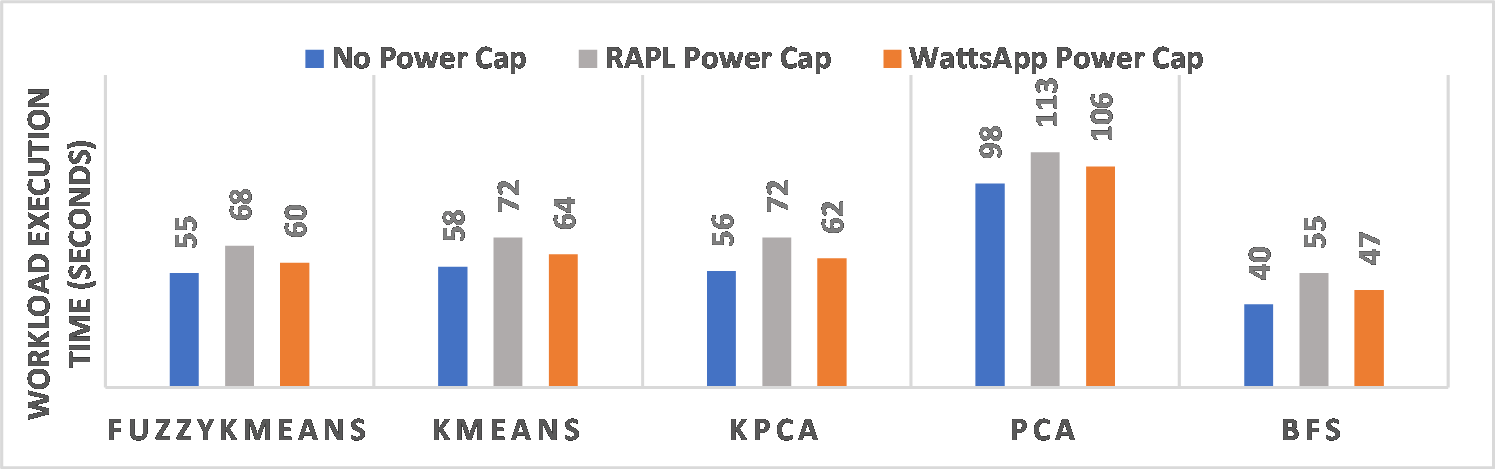}}
\\
	\subfloat[For MPI-C programs from Table~\ref{table:workloads}]
	{\label{fig:execution-time-single-2}
	\includegraphics[width=0.48\textwidth]
	{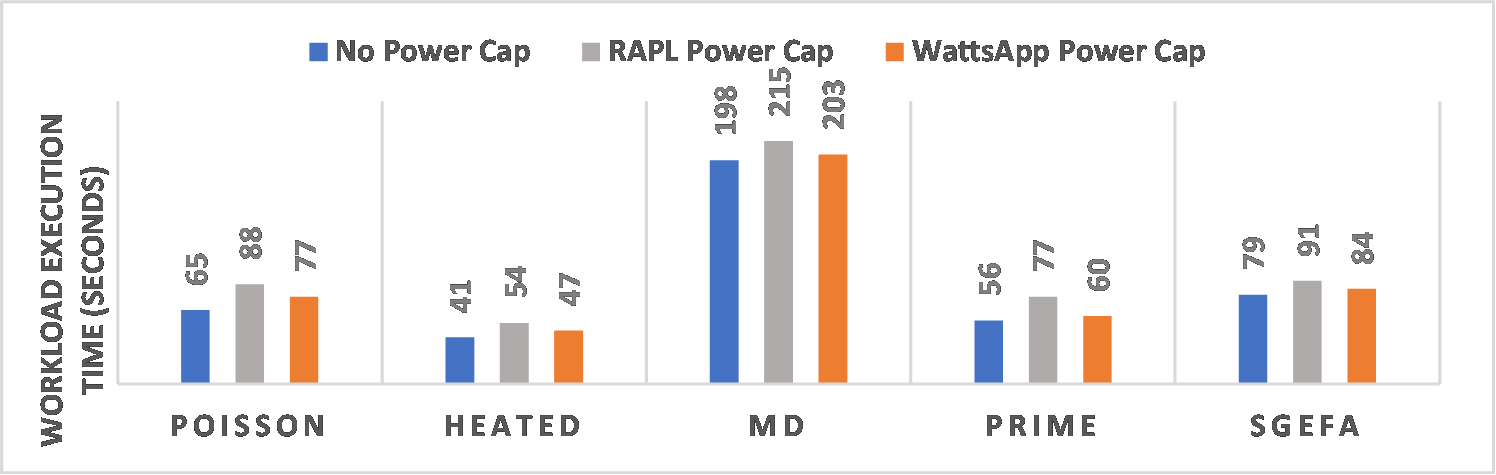}}
\end{center}
\caption{Execution time of individual workloads when a single container executes on a server using the power cap technique on the Intel Xeon processor}
\label{fig:execution-time-single}
\end{figure}

Figure~\ref{fig:execution-time-single} shows the results for the first experiment in which a single container executes on the server with a given workload. The graph shows the workload execution time for the proposed power cap method, no power cap, and RAPL's power cap is adopted. In this case, it is noted that the proposed power capping method is more effective than RAPL's power cap since the total workload execution time is lower in every case. This highlights the benefit of using the proposed power-aware container scheduling method. 

\begin{figure}[t]
\begin{center}
	\subfloat[For DCBench programs from Table~\ref{table:workloads}]
	{\label{fig:execution-time-multi-1}
	\includegraphics[width=0.48\textwidth]
	{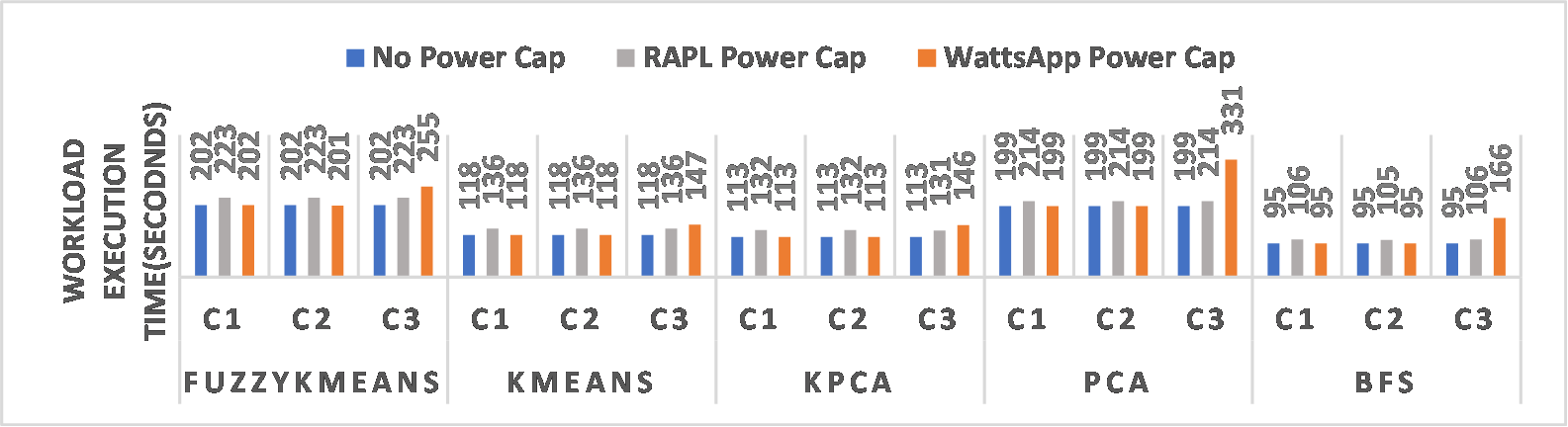}}
\\
	\subfloat[For MPI-C programs from Table~\ref{table:workloads}]
	{\label{fig:execution-time-multi-2}
	\includegraphics[width=0.48\textwidth]
	{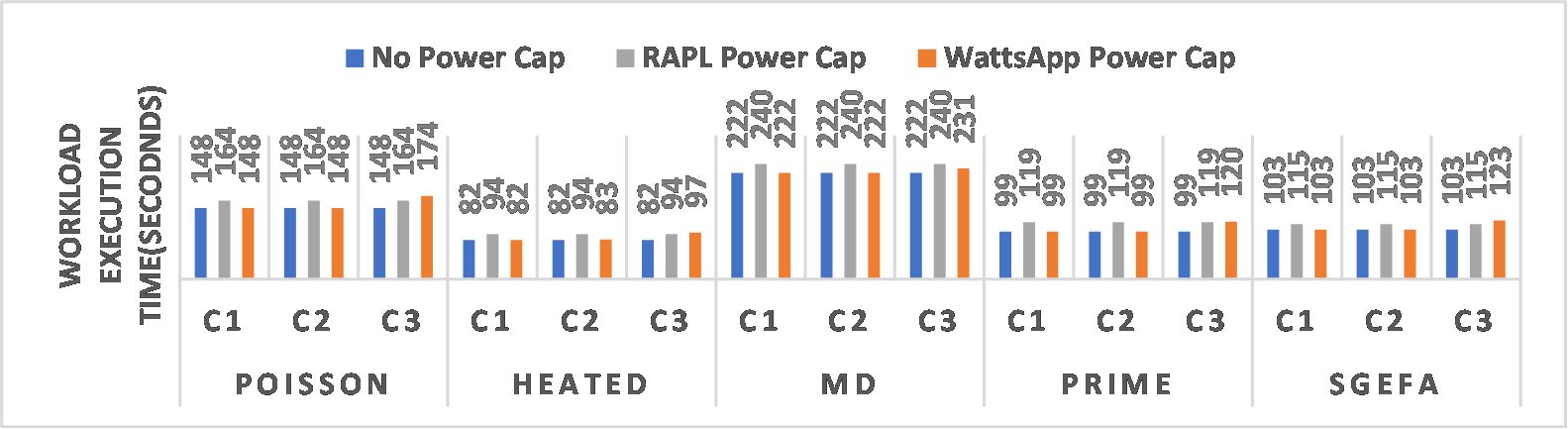}}
\end{center} 
\caption{Execution time of individual workloads in relation to the power cap technique when multiple containers execute on the same server on the Intel Xeon processor}
\label{fig:execution-time-muliti}
\end{figure}

Figure~\ref{fig:execution-time-muliti} shows the results for the second experiment in which three containers (C1, C2, C3) with the same workloads execute on the server. These workloads are more representative of a real world scenario. The graph shows the workload execution time when the proposed power cap method, no power cap, and RAPL's power cap is adopted. It can be clearly seen that when RAPL is employed the workload execution time of all containers increase. This is because RAPL achieves power capping by reducing the CPU frequency of the server, which in turn affects the performance of all containers running on the server. On the other hand, it is noted that the proposed power cap technique reduces the allocated number of CPU cores, thus degrading the performance of only one container as opposed to all the running containers. Therefore, `C3' in many cases is noted to take longer than the other containers. In short, only one container out of many (that potentially violates the power cap) is penalized when using the proposed power cap technique. 

\begin{figure}[t]
  \centering
  \includegraphics[height=80pt,width=0.48\textwidth]{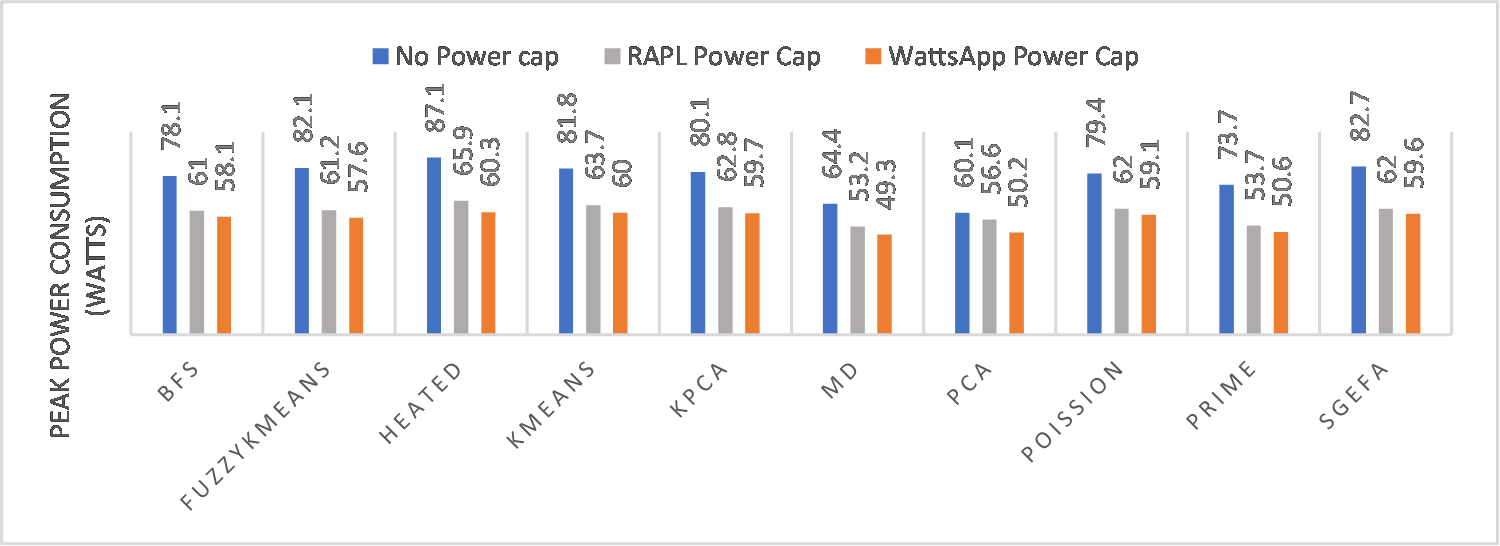}
  \caption{Peak power consumption of the workloads under the power cap techniques on the Intel Xeon processor}
  \label{fig:peakpower}
\end{figure}

Figure~\ref{fig:peakpower} shows the peak power consumption on the Intel Xeon processor of applications (from Table~\ref{table:workloads}) when there is no power capping, under the \texttt{WattsApp} power capping regime and the RAPL power capping technique. This experiment uses the power cap limit of 55W. The average peak power consumption of the proposed power capping technique is 56.4W which is close to the power cap limit where as the average peak power of RAPL's power cap is 60.2W and significantly higher than the power capping limit. The peak power consumption for \texttt{WattsApp} is 60.3W in comparison to the peak power consumption of RAPL's power cap is 65.9W.

Similar experiments are carried out on the Odroid board to demonstrate the effectiveness of \texttt{WattsApp} power capping on ARM processors. Again only six workloads from Table~\ref{table:workloads} are used. As RAPL is specific to Intel, these experiments only compare the \texttt{WattsApp} power cap with no power cap.

\begin{figure}[t]
  \centering
  \includegraphics[width=0.48\textwidth]{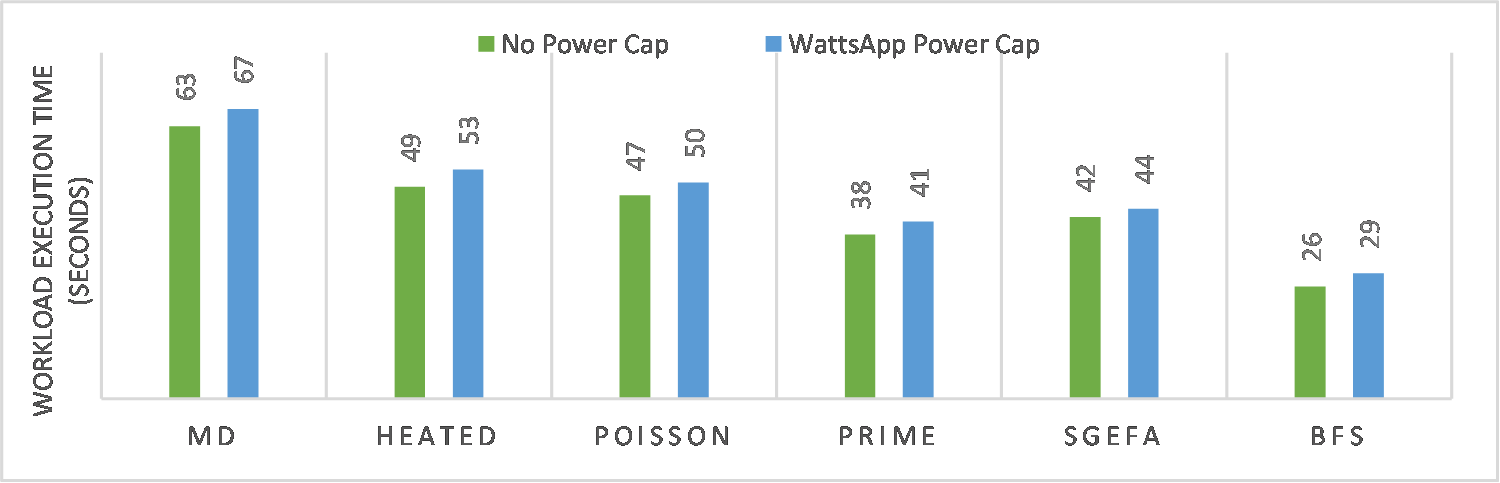}
  \caption{Execution time of individual workloads when a single container executes on a server using the power cap technique on ARM processor}
  \label{fig:odroid-single}
\end{figure}

Figure~\ref{fig:odroid-single} shows the results of the experiment in which a single container executes on the ARM processor. The workloads running under \texttt{WattsApp} power cap takes slightly longer time and executes within the power budget defined by power cap limit of 7W.

\begin{figure}[t]
  \centering
  \includegraphics[width=0.48\textwidth]{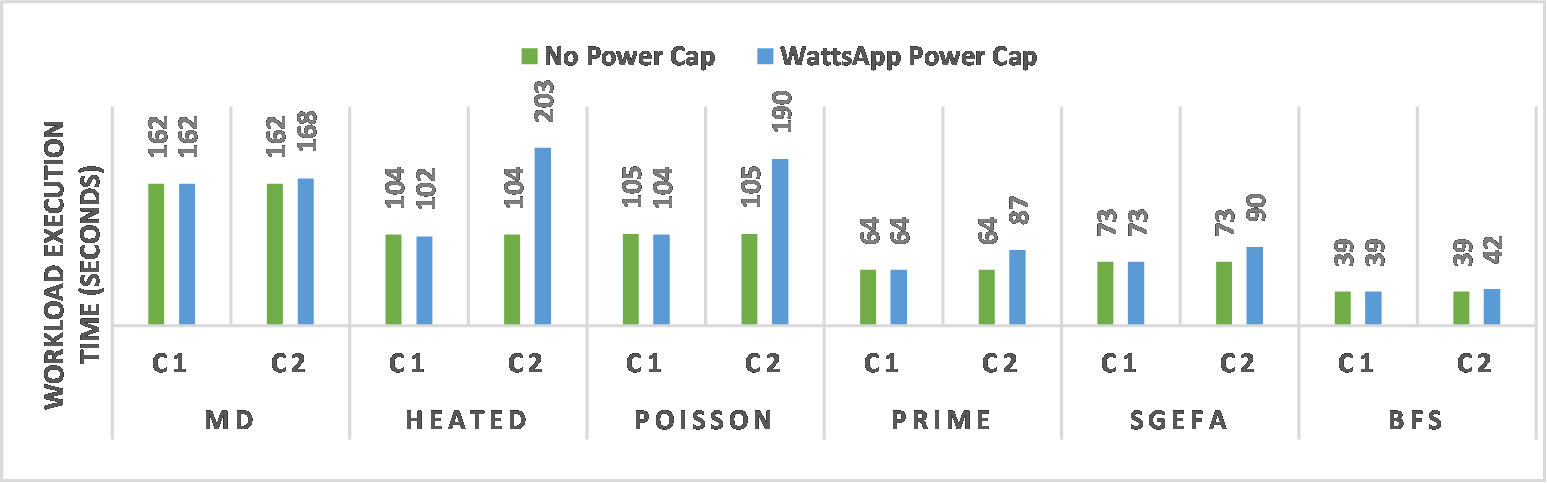}
  \caption{Execution time of individual workloads in relation to the power cap technique when multiple containers execute on the same ARM processor}
  \label{fig:odroid-multi}
\end{figure}

Figure~\ref{fig:odroid-multi} shows the results when two containers with the same workload executes on the ARM processor. The graph shows the workload execution time under the no power cap and \texttt{WattsApp} power cap. Again workloads running under \texttt{WattsApp} power cap takes longer time and the power budget remains below the power cap limit of 9W. This demonstrate that \texttt{WattsApp} power capping in also effective on ARM processors.

\begin{figure}[t]
  \centering
  \includegraphics[width=0.48\textwidth]{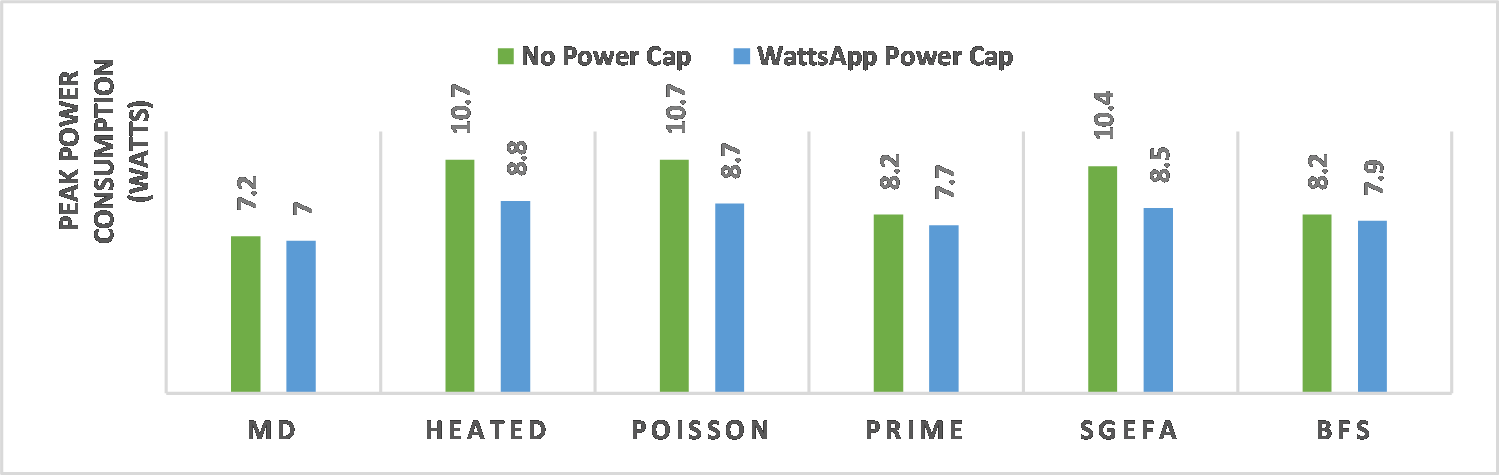}
  \caption{Peak power consumption of the workloads under the power cap techniques on the ARM processor}
  \label{fig:peakpowerodroid}
\end{figure}

Figure~\ref{fig:peakpowerodroid} shows the peak power consumption of the six workloads from Table~\ref{table:workloads} when there is no power capping and under the \texttt{WattsApp} power capping regime. This experiment uses the power cap limit of 9W. The average peak power of the proposed power capping technique is 8.1W (below the power cap limit), but is 9.2W when there is no power cap. 

\begin{figure}[t]
  \centering
  \includegraphics[width=0.48\textwidth]{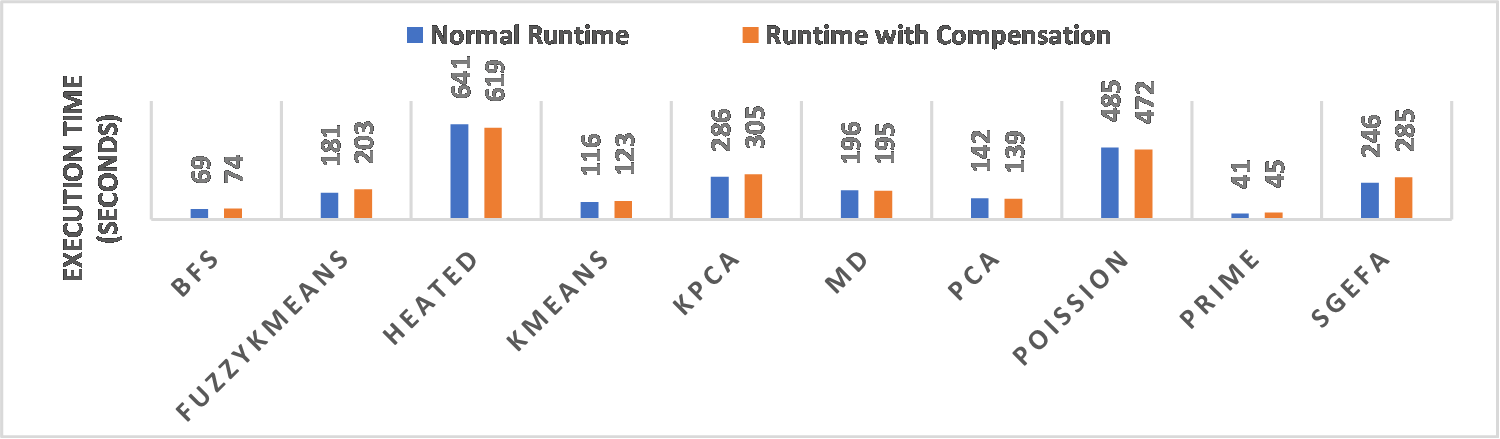}
  \caption{Effect of power capping with compensation on parallel workloads executing on the Intel Xeon processor}
  \label{fig:compensation}
\end{figure}

\begin{figure}[t]
  \centering
  \includegraphics[width=0.48\textwidth]{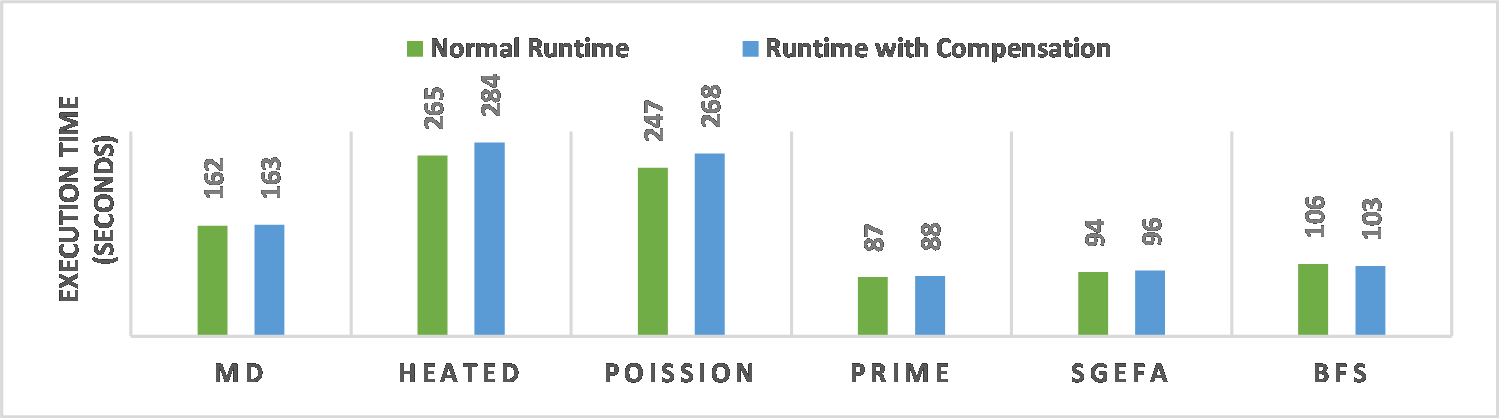}
  \caption{Effect of power capping with compensation on parallel workloads executing on the ARM processor}
  \label{fig:compensationarm}
\end{figure}

Figure~\ref{fig:compensation} and Figure~\ref{fig:compensationarm} demonstrates another aspect - when a single MPI application is executed across a cluster of containers. This experiment considers that workloads are running on a cluster of two or more different servers. 
 When the CPU cores of the workload need to be reduced on one server, then it is compensated for by increasing the CPU cores allocated on the other server for the workload.
The result shows that power capping on one server with compensatory allocation on another server does not significantly impact performance. 

\textbf{\textit{Summary:}}
In short, the experimental results highlight that:
(i) The data collection overhead in the proposed power-aware container scheduling method of \texttt{WattsApp} only affects the system power consumption negligibly.
(ii) Nearly 90\% of data samples are estimated with less than 10\% error.
(iii) The MAPE of power estimation using the model that is employed by the proposed power-aware container scheduling method of \texttt{WattsApp} is between 1\%-6\%. This is relatively low and accurate estimations can be expected from the model. 
(iv) The power estimation method of \texttt{WattsApp} is also validated on the parallel applications across a cluster of containers. The proposed model estimate the power consumption with MAPE between 1\% to 5.5\%. The impact of power capping (CPU core reduction) for parallel workloads is minimized during the workload runtime by applying compensation.
(v) Deallocation of resources are found to be a more feasible approach than migrating containers for the power capping technique given that the overheads for migration increase with the size of the container. The overheads for deallocating resources is negligible. 
(vi) When both single and multiple containers are executed on the server, it is noted that the proposed power cap method is more beneficial than when no power cap or RAPL's power cap is employed since the proposed method does not degrade the performance of all running containers to keep the power budget below the cap. \texttt{WattsApp}'s power capping is also effective since the peak power allowed by the \texttt{WattsApp} method is less than that of RAPL's power cap and does not violate the soft power cap that may be imposed by administrators.

\section{Related Work}
\label{sec:relatedwork}
Power modeling of processors and VMs are well explored, but power modeling of containers is still in its early stages.
In this section, the impact of virtualization techniques on system power consumption that do not use estimation models is discussed. Then container power modeling techniques and power models for processors/servers and VMs are considered. Finally, research on container power capping is discussed.

\textbf{Impact of virtualization techniques on system power}: Power consumed by servers running containers has been experimentally measured without developing power models. The CPU usage of the container has a significant contribution to the overall power consumption~\cite{tadesse2017energy}. 
Empirical investigation on four virtualization technologies, namely Xen, KVM, Docker and LXC is noted~\cite{morabito2015power}. It was observed that for CPU workloads there is no significant difference in power consumption among the above technologies. However, containers consume less power than other virtualization technologies for network-based workloads. 

A comparison of the server power consumption~\cite{van2016power} and energy comparison~\cite{jiang2017energy} has been presented for virtualization and containerization technologies. The power and energy characteristics of four hypervisors and a container engine including VMware ESXi, Microsoft Hyper-V, KVM, XenServer and Docker on six different hardware (three mainstream 2U rack servers, one emerging ARM64 server, one desktop server, and one laptop) is considered. It is observed that hypervisors exhibit different power and energy consumption profiles when the same workload is executed on the hardware. Although containers are light weight, they are not necessarily more power-efficient than VMs.
Similar comparisons of running workloads on containers and bare-metal servers are considered~\cite{santos2018does}. It is observed that running Docker has an inherent power cost and thus energy consumed is higher than bare-metal.

\textbf{Container Power Modeling}: SmartWatts~\cite{Fieni2020} is a self calibrating software power model for containers that relies on hardware performance counters and RAPL's power measurement of CPU and DRAM for estimating power. Using RAPL limits the applicability of SmartWatts to Intel architectures. It also does not capture the impact of disk access and network usage on power that may be the main activity of an I/O or a network based application. The power model of \texttt{WattsApp} on the other hand uses architecture agnostic parameters to model system power and its feasibility on multiple hardware platforms is demonstrated. 

Lightweight power models, such as cWatts+~\cite{phung2017application} and cWatts++~\cite{phung2019lightweight} are developed for containers. cWatts++ is a virtual power model that has two components: a client back-end and a server front-end. The client back-end is installed in the container and accesses the CPU event counters. cWatts++ uses two models, namely an event-based and RAPL-based models. The event-based model uses CPU performance counters and RAPL-based models uses only RAPL event counters.
The evaluation shows that the two power models are useful on workloads obtained from the PARSEC and in-house benchmarks. However, cWatts only uses CPU related metrics to compute container power from server power. CPUs are a major power consuming component of a typical server (nearly one-third~\cite{colmant2015} and even up to 40\%~\cite{mccullough2011evaluating} of the total server power), but other components need to be considered. Hence, \texttt{WattsApp} considers memory, IO and the network to account for container power. Moreover, cWatts is intrusive and requires client installations and access to containers (this may not be always feasible depending on access permissions, ownership models, business models etc). 

There is research that accounts for the power consumption of individual threads and application containers~\cite{brondolin2018deep}. The research relies on power estimation of each CPU core obtained from Intel's RAPL and hardware performance counters (related to CPU events) obtained from the OS.
A power-aware consolidation technique of containerized data centers based on a model built using CPU utilization is presented~\cite{khan2019energy}.
Both approaches are based on CPU related metrics and do not account for the power consumed by other components~\cite{mccullough2011evaluating}). \texttt{WattsApp} on the other hand considers CPU, memory, disk and network related metrics.

\textbf{Power Models for VM and Processor Power Models}: Joulemeter~\cite{kansal2010} is a software power meter for VMs. Joulemeter estimates VM power using linear regression with ordinary least square to correlate VM profiling data to system power consumption.

BITWATTS uses a two-level approach to estimate the process level power estimation based on CPU power consumption is developed~\cite{colmant2015}. This approach performs profiling at both system and VM level. The system-level profiling estimates the system-level power consumption whereas, the VM level profiling estimates the power consumption of hosted applications. BITWATTS uses a regression technique for learning the process level power models.

Another approach uses a tree regression-based technique for VM power metering~\cite{gu2015}. It is suggested that the linear regression methods are not sufficiently accurate and therefore an approach that recursively partitions the collected data into easy modeling pieces has been proposed. This method first, builds the server model using the observation for server and then fairly divides this server power consumption among the virtual machines.

iMeter is a performance counters based VM power model based on polynomial kernel based support vector regression~\cite{yang2014}. Principal component analysis is performed to select performance counters that have the most impact on power consumption.

Similarly, there are power models for processors and the components of a server. A two stage cross architecture power model is discussed in~\cite{chen2020cross}. This model takes advantage of both
linear regression and support vector machines to provide power estimation with a high accuracy across multiple hardware architectures. 

A configurable learning framework named PowerAPI that automatically builds the power model of a CPU is presented~\cite{colmant2018next}. PowerAPI automatically explores the available hardware performance counters of a CPU and selects the performance counter having the most impact on the power consumption of the server. 

There are a number of other notable processor power consumption models proposed by researchers (for example,~\cite{bircher2005runtime, bertran2010decomposable, bertran2012counter, dargie2015stochastic}), which are not considered within the scope of the discussion of this paper. 

However, \texttt{WattsApp} is a software power model that uses container resource utilization statistics obtained from the host OS to estimate the container power consumption.

\textbf{Container Power Capping:} There is limited research in the literature that focuses on power capping for containers. Two power capping techniques are proposed in literature:

The first is a power capping technique (DockerCap) for Docker containers~\cite{asnaghi2016dockercap}. The system power consumption is obtained from the hardware power meter and RAPL. The CPU quota of all the containers of different priority is reduced, thereby affecting the performance of all the containers. 
The \texttt{WattsApp} method however uses two techniques, namely container migration, and CPU core reduction to achieve power capping. The merit of the proposed method is that the overall container performance is unaffected and only the container that violates the power cap is degraded.

The second power capping technique is proposed for Docker containers on the Kubernetes platform~\cite{arnaboldi2018hyppo} by relying on DEEP-mon power monitoring ~\cite{brondolin2018deep}. This technique relies on RAPL and DVFS to manage power cap limits. 
It is demonstrated in this paper that using RAPL affects the run-time performance of all containers running on the server. RAPL enforces a power cap on the processor package and DRAM by reducing the CPU frequency and thus degrades the overall system performance. However, \texttt{WattsApp} uses architecture independent metrics to measure resource utilization (CPU, memory, disk and network) making the approach portable. The proposed method is demonstrated to be effective for both power-aware scheduling and capping.

\section{Conclusions}
\label{sec:conclusions}
This paper proposes \texttt{WattsApp} that is underpinned by a six-step power-aware scheduling method for containers to minimize power cap violations on a server in real-time. The method relies on a neural network-based power estimation model. The trained model effectively predicts over 90\% of data samples with less than 10\% error. By testing on 10 representative benchmark workloads, the approach is able to achieve a MAPE error of less than 6\%, and displays minimal overhead during run time scheduling. Unlike hardware-based power capping techniques, such as Intel's RAPL, which are indiscriminate to workloads and degrade the overall performance of all containers running on a server, this software-based approach is able to target individual containers running workloads, minimizing overall processing degradation while maintaining a node's power budget. The proposed technique considers multiple scenarios, including (i) single/multiple application, single container and single application, multiple containers. \texttt{WattsApp} has been shown to be feasible and outperforms existing techniques.

\textbf{\textit{Future Work:}}
\texttt{WattsApp} is tested on two popular processor types. However, to deal with heterogeneous platforms the power model will need to be expanded further to accelerator architectures. 

The current method prioritizes the power budget of an individual server, but not the performance or SLA of the container workloads. Better understanding, profiling, or feedback (for example, as presented in Barrelfish~\cite{Baumann:2009:MNO:1629575.1629579}) from an application would enable better choices within the scheduler and is an avenue for future work. 

Finally, this technique will be applied to edge computing, where power is a critical concern. As containers are being increasingly used in this space, this research is directly applicable. Alternate types of workloads, such as stream processing and sensor-based applications will be considered.

\begin{acks}
Blesson Varghese is supported by a Royal Society Short Industry Fellowship and by funds from Rakuten Mobile, Japan.
\end{acks}

\bibliographystyle{ACM-Reference-Format}
\bibliography{reference}

\end{document}